\documentclass{article} 
\usepackage{iclr2026_conference}

\usepackage{amsmath,amsfonts,bm}

\def\eqref#1{equation~\ref{#1}}

\def\1{\bm{1}}

\DeclareMathAlphabet{\mathsfit}{\encodingdefault}{\sfdefault}{m}{sl}
\SetMathAlphabet{\mathsfit}{bold}{\encodingdefault}{\sfdefault}{bx}{n}

\usepackage{graphicx}
\usepackage{hyperref}
\usepackage{url}
\usepackage{lipsum}
\usepackage{bm} 
\usepackage{amsmath}
\usepackage{graphicx} 
\usepackage{cleveref}
\usepackage{booktabs}
\usepackage{multirow}
\title{$\mathcal{X}$-Part: high fidelity and structure coherent shape decomposition}

\vspace{-5cm}
\author{\small Xinhao Yan,$^{1,2,}\thanks{Equal Contributions}$, Jiachen Xu,$^{1,*}$, Yang Li,$^{1,\thanks{Project Leader}}$, Changfeng Ma$^{1,3}$, Yunhan Yang$^{1,4}$, 
\And  \small \hspace{-0.7cm} Chunshi Wang$^{1,5}$, Zibo Zhao$^1$, Zeqiang Lai$^{1,6}$, Yunfei Zhao$^{1}$, Zhuo Chen$^{1}$, Chunchao Guo$^{1, \thanks{Corresponding Author}}$ 
\vspace{0.3cm}
\And
\normalfont \hspace{1cm} $^1$ Tencent Hunyuan, $^2$ShanghaiTech, $^3$NJU,  $^4$HKU, $^5$ZJU, $^6$CUHK  \\ 
\vspace{-25cm}
}

\iclrfinalcopy
\begin{document}

\maketitle

\begin{figure}[!h]
    \centering
    \includegraphics[width=1\linewidth]{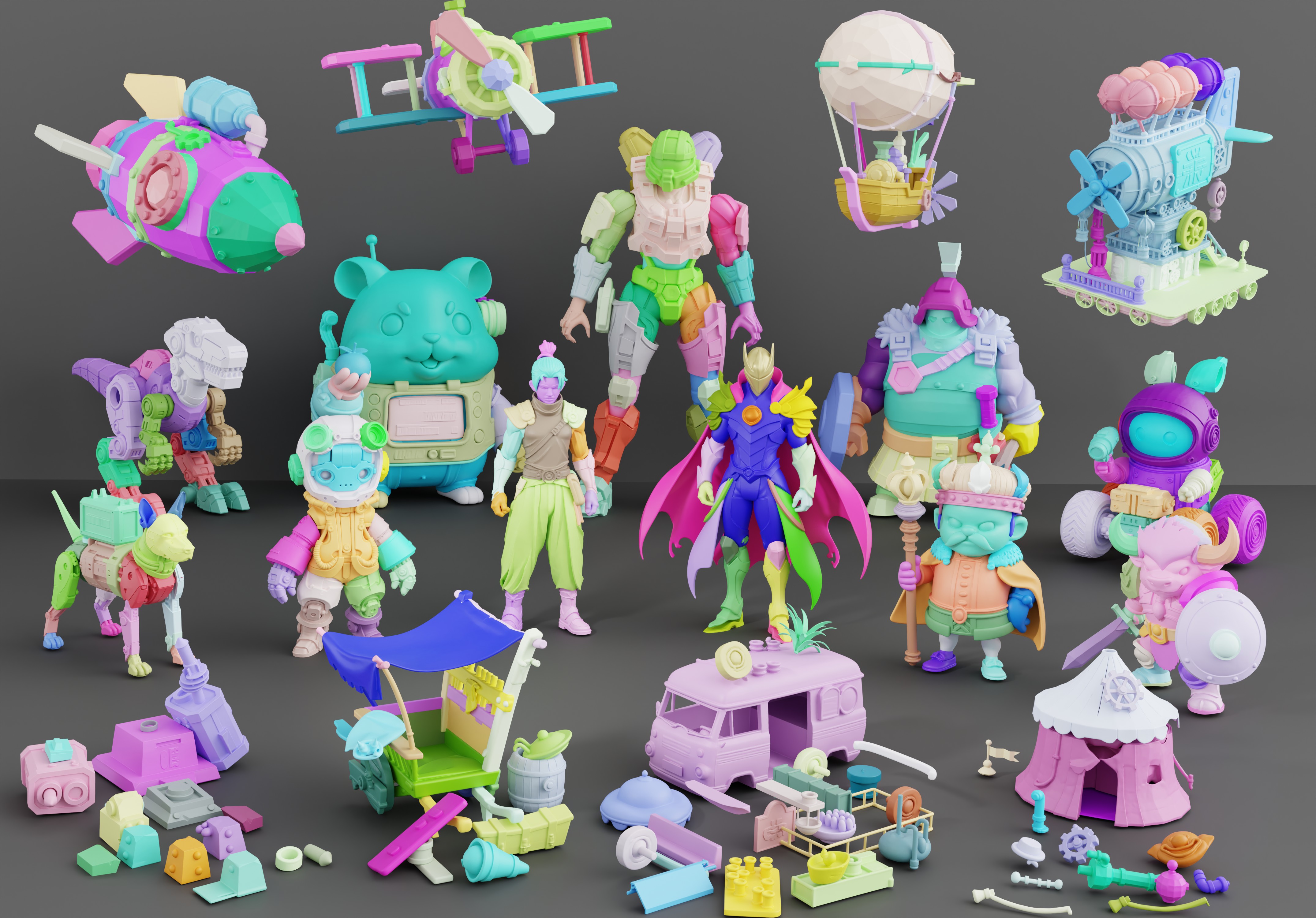}
    \label{fig:teaser}
\end{figure}
\begin{abstract}

Generating 3D shapes at part level is pivotal for downstream applications such as mesh retopology, UV mapping, and 3D printing.
However, existing part-based generation methods often lack sufficient controllability and suffer from poor semantically meaningful decomposition.
To this end, we introduce $\mathcal{X}$-Part, a controllable generative model designed to decompose a holistic 3D object into semantically meaningful and structurally coherent parts with high geometric fidelity. $\mathcal{X}$-Part exploits the bounding box as prompts for the part generation and injects point-wise semantic features for meaningful decomposition. Furthermore, we design an editable pipeline for interactive part generation.
Extensive experimental results show that $\mathcal{X}$-Part achieves state-of-the-art performance in part-level shape generation. 
This work establishes a new paradigm for creating production-ready, editable, and structurally sound 3D assets. Codes will be released for public research.
\end{abstract}

\section{Introduction}

3D assets are now extensively utilized across a wide range of fields, including gaming, film production, 3D printing, autonomous driving, and robotic simulation.
However, traditional 3D content creation remains a time-consuming process that demands significant expertise. 
Recent advances in generative AI have substantially lowered the barriers to 3D content generation particularly with the emergence of foundational 3D models~\cite{zhang2024clay,zhao2025hunyuan3d,lai2025hunyuan3d}. 

Despite this progress, most existing generative approaches are only capable of producing monolithic 3D models, which poses considerable limitations for practical 3D creation pipelines. 
Decomposing a complete 3D shape into meaningful semantic parts would greatly facilitate various downstream tasks. 
For instance, breaking down a complex geometry into simpler parts can significantly ease the process of mesh re-topology~\cite{weng2025scaling} and uv-unwrapping~\cite{li2025auto}.
Generating shapes at the part level presents two major challenges: 1) The decomposed geometry must maintain meaningful part-level semantics, and 2) The generation process must recover geometrically plausible structures for internal regions.

Mainstream part-generation methods adopt the latent vecset diffusion framework~\cite{zhang20233dshape2vecset}, where each part is represented as an independent set of latent codes for diffusion. 
The generation process can be executed independently for individual parts (e.g., HoloPart~\cite{yang2025holopart}) or simultaneously for all parts (e.g., PartCrafter~\cite{lin2025partcrafter}, PartPacker~\cite{tang2024partpacker}) with enhanced part synchronization. Furthermore, multi-view or 3D segmentation are frequently employed for better part decomposition~\cite{chen2024partgen,yang2025holopart,yang2025omnipart}. However, these approaches are highly sensitive to inaccuracies in the segmentation results. Alternative works~\cite{lin2025partcrafter,tang2024partpacker} do not explicitly rely on segmentation, but they lack controllability and often generate parts with ambiguous boundary.

Motivated by these observations, we present  $\mathcal{X}$-Part, a diffusion-based framework that decomposes a holistic mesh into semantically meaningful and structurally coherent 3D parts.
The method utilizes the state-of-the-art segmenter P$^3$-SAM~\cite{P3_SAM} to automatically generate initial part segmentations, bounding boxes, and semantic features. Then the shape decomposition is executed within a synchronized multi-part diffusion process.

Specifically, 1) First, to control part decomposition, 
instead of directly using segmentation results as input
we uses bounding boxes as prompts to indicate part locations and scales.
Compared with fine-grained and point-level segmentation cues, bounding boxes provide a coarser form of guidance, which mitigates overfitting to the input segmentation masks. Besides, the bounding box provides additional volume scale information for the partially visible part, benefiting generation and controllability.
2) Second, despite inaccuracies in the segmentation results, we notice that the high-dimension point-wise semantic feature is free from the information compression caused by the mask prediction head used in P$^3$-SAM, resulting in more robust semantic representations. Therefore, we introduce the semantic features from P$^3$-SAM into our diffusion process to guide the multi-part diffusion process. This greatly benefits the part decomposition. 
3) Third, we integrate $\mathcal{X}$-Part into a bounding box based part editing pipeline following~\cite{lugmayr2023inpainting}. It supports local editing, such as splitting a part into several parts and adjusting their scales, to facilitate interactive part generation.

To prove the effectiveness of $\mathcal{X}$-Part, we conducted extensive experiments on various benchmarks. Our results show that $\mathcal{X}$-Part achieves state-of-the-art performance in part-level decomposition and generation. In summary, the contributions of our work are as follows:
\begin{enumerate}
\item We propose $\mathcal{X}$-Part, a controllable and editable diffusion framework, capable of generating semantically meaningful and structurally coherent 3D parts.
\item We integrate $\mathcal{X}$-Part into an editable part generation pipeline, which supports multiple interactive editing methods.
\item Extensive experiments demonstrate that $\mathcal{X}$-Part achieves state-of-the-art performance in part-level decomposition and generation.
\end{enumerate}

\section{Related Work}
\textbf{Part Segmentation.} The most straightforward approach for decomposing a 3D geometry is segmentation. Early conventional methods~\cite{qi2017pointnet,zhao2021point} directly predict per-point semantic labels via supervised learning. While effective within constrained settings, these methods rely heavily on extensive part-level annotations, generalize poorly beyond seen categories, and offer limited semantic scalability. Inspired by the remarkable success of 2D foundation models like SAM~\cite{kirillov2023segment} and GLIP~\cite{li2022grounded} in open-vocabulary tasks, several recent approaches~\cite{abdelreheem2023satr,liu2023partslip,tang2024segment,thai20243,umam2024partdistill,zhong2024meshsegmenter} attempt to lift 2D visual knowledge to 3D domains. Although these methods improve generalization, they often fail to accurately infer parts in occluded or unobserved regions. To mitigate this, PartField~\cite{liu2025partfield} and SAMPart3D~\cite{yang2024sampart3d} learn open-world 3D feature fields for semantic part decomposition. 
P3-SAM~\cite{P3_SAM} proposes a native 3D part segmentation network trained on a large, purely 3D dataset with part annotations, demonstrating impressive part segmentation results.

\noindent\textbf{Object-level Shape Generation.}
The remarkable success of latent diffusion models in 2D image generation has inspired a new wave of methods extending this capability to 3D object generation. Dreamfusion~\cite{poole2022dreamfusion} introduced Score Distillation Sampling (SDS) to distill 2D priors from pre-trained diffusion models for 3D synthesis, though it often suffers from slow optimization and geometrically inconsistent outputs. Subsequent approaches~\cite{li2023instant3d,long2024wonder3d,shi2023mvdream}, reformulated 3D generation as a multi-view image synthesis problem. With the release of large-scale 3D datasets such as Objaverse~\cite{deitke2023objaverse} and Objaverse-XL~\cite{deitke2023objaversexl}, native 3D generative models have become increasingly prevalent. Methods like 3DShape2VecSet~\cite{zhang20233dshape2vecset}, Michelangelo~\cite{zhao2023michelangelo}, Clay~\cite{zhang2024clay}, and Dora~\cite{chen2025dora} encode object point clouds into vector-set tokens using a variational autoencoder (VAE)~\cite{kingma2013auto} and model the distribution via a Diffusion Transformer (DiT)~\cite{peebles2023scalable}. In contrast, Trellis~\cite{xiang2025structured} employs an explicit voxel representation for coarse geometry and further generates both geometry and appearance from the voxel latents.

\noindent\textbf{Part-level Shape Generation.}
PartGen~\cite{chen2025partgen} decomposes 3D objects by solving a multi-view segmentation task and subsequently completes and reconstructs each part in 3D. PhyCAGE~\cite{yan2024phycage} adopt physical regularization for non-rigid part decomposition. The second category exploits DiT-based generative methods to achieve part-level generation~\cite{yang2025holopart,lin2025partcrafter,tang2024partpacker,dong2025one,yang2025omnipart,zhang2025bang}. HoloPart~\cite{yang2025holopart} completes part geometry from initial 3D segmentation results. In contrast, PartCrafter~\cite{lin2025partcrafter} and PartPacker~\cite{tang2024partpacker} operate without explicit segmentation, instead leveraging multi-instance DiTs to generate parts automatically. PartPacker~\cite{tang2024partpacker} further introduces a dual-volume DiT to model complementary spatial volumes for improved efficiency. 
Frankenstein~\cite{yan2024frankenstein} execute similar idea by packing multiple SDFs in a latent triplane space via VAE.
However, these approaches often yield parts with limited geometric quality and offer minimal local controllability. 
CoPart~\cite{dong2025one} incorporates an auxiliary 2D image diffusion model to enhance texture and detail using 2D/3D bounding box conditions, though it supports only up to 8 parts and cannot decompose an existing 3D shape. OmniPart~\cite{yang2025omnipart} adopts an explicit representation similar to Trellis and uses bounding box prompts, yet it lacks the ability to complete occluded geometry. BANG~\cite{zhang2025bang} frames part generation as an object explosion process, enabling bounding-box-guided decomposition and recursive refinement, but it often fails to preserve fine geometric details throughout the process. AutoPartGen~\cite{chen2025autopartgen} employs a latent diffusion model to autoregressively generate parts, which is computationally expensive and offers limited user control.
Kestrel~\cite{ahmed2024kestrel} employs an LLM to reason about part segmentation, while ShapeLLM~\cite{qi2024shapellm} leverages LLMs for visual grounding and bounding box extraction.
MeshCoder~\cite{dai2025meshcoder} represents parts using a code-based representation, injects 3D point clouds into an LLM, and generates 3D primitives expressed as Blender Python scripts.

\section{Method}

\begin{figure*}
  \centering
  	\includegraphics[width=1.\linewidth]{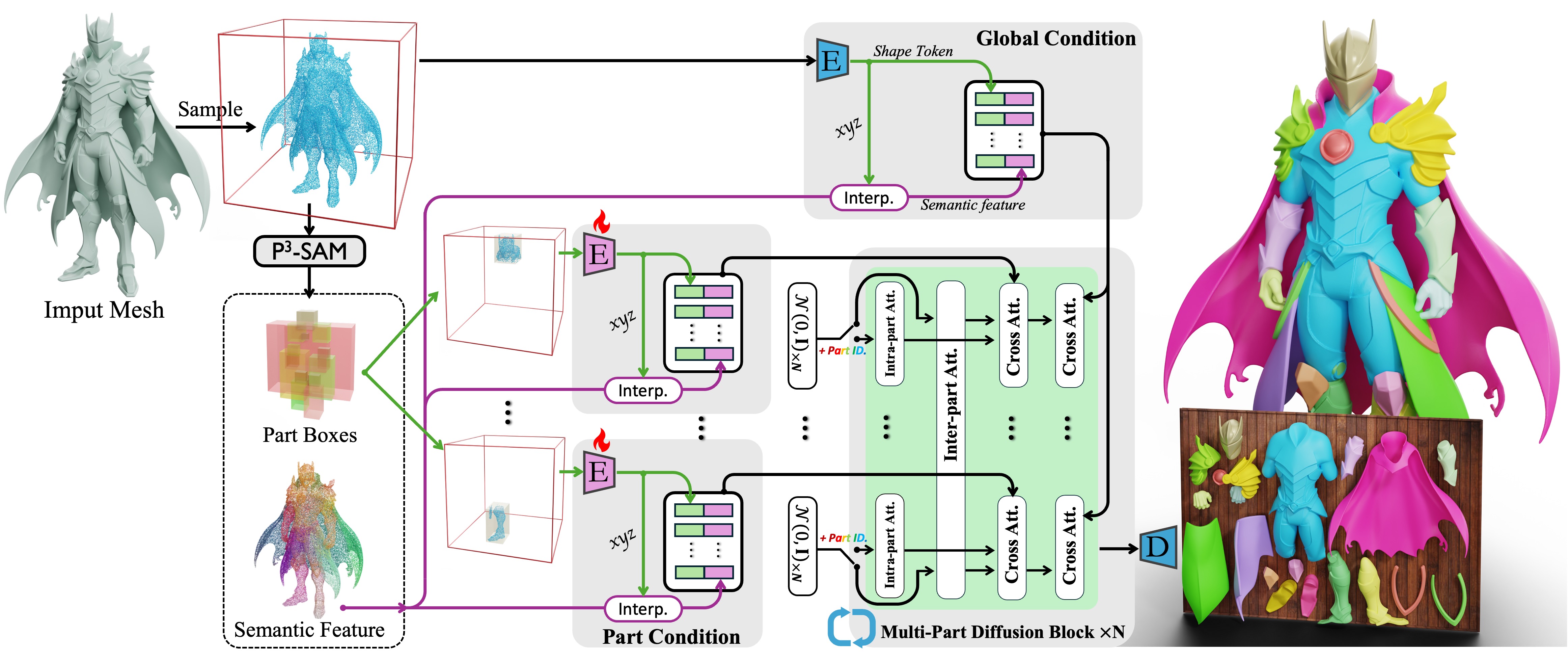}
  \caption{\textbf{Architecture of $\mathcal{X}$-Part}. 
  Given input point cloud, per-point feature and part bounding boxes are extracted from $\text{P}^3\text{-SAM}$.
  Global and part conditions are obtained by stacking geometry token with interpolated semantic features. 
  They are injected to multi-part diffusion process to guide shape decomposition.
  }
  \label{fig:pipeline}
\end{figure*}

Our objective is to generate high-fidelity and structure-coherent part geometries from a given object point cloud, while ensuring flexible controllability over the decomposition process. To this end, we propose $\mathcal{X}$-part (see \Cref{fig:pipeline}) based on the diffusion framework. In \Cref{sec:preliminary}, we outline the foundational vecset-based 3D latent diffusion framework. In \Cref{sec:stage_1} we first describe our proposed bounding box-based part-level cues extraction module and point-wise semantic feature injection, then we present our overall $\mathcal{X}$-Part framework for  synchronized part generation and training scheme. Finally, we introduce the editable part generation pipeline in \Cref{sec:edit_pipeline}.

\subsection{Preliminary \label{sec:preliminary}}
Our method builds upon pre-trained vecset-based 3D shape generation models \cite{zhang2024clay,zhao2023michelangelo,zhao2025hunyuan3d,li2025triposg}, which typically consist of a 3D shape variational autoencoder (VAE) and a latent diffusion model. 

\noindent\textbf{Variational Autoencoder (VAE)} Given an input mesh, we first sample a point cloud $\bm{X} \in \mathbb{R}^{N\times 7}$ with both normal and sharp-edge flags using a sharp-edge-aware sampling strategy, following Dora \cite{chen2025dora} and Hunyuan2.0 \cite{zhao2025hunyuan3d}. The VAE adopts a transformer-based architecture \cite{zhang20233dshape2vecset, zhang2024clay,li2025triposg,zhao2025hunyuan3d}, comprising a cross-attention block followed by multiple self-attention layers, whose encoder maps the sampled point cloud into latent vectors:
\begin{equation}
    \bm{Z}=\mathcal{E}(\bm{X})= \text{SelfAttn}( \text{CrossAttn} (PE(\bm{X_0}),PE(\bm{X})))
\end{equation}
where $\bm{X_0} \in \mathbb{R}^{N_0\times 7}$ denotes the point set obtained by applying farthest point sampling (FPS) to $\bm{X}$, and $\bm{Z}\in \mathbb{R}^{N_0\times C}$ represents the $N_0$ latent tokens of the input shape. $PE$ represents position embedding for input point cloud.

The decoder of the VAE similarly consists of several self-attention layers followed by a final cross-attention module, mapping a spatial coordinate query $q\in\mathbb{R}^{3}$ to its corresponding signed distance value (SDF):
\begin{equation}
    \mathcal{D}(q, \bm{Z})=\text{CrossAttn}(q, \text{SelfAttn}(\bm{Z}))
\end{equation}
Note that, to enhance the capacity of VAE to represent part-level geometry, we further fine-tune the VAE on the part-level dataset. 

\noindent\textbf{3D Diffusion Model} To model the latent space of encoded objects, a flow-based diffusion model~\cite{lipman2022flow} is trained to generate latent tokens, which can subsequently be decoded into 3D geometries. Following Hunyuan-DiT \cite{li2024hunyuan} and TripoSG \cite{li2025triposg}, the core of our model is constructed using a series of Diffusion Transformer (DiT) blocks.

\subsection{Multi-Parts Latent Diffusion \label{sec:stage_1}}
\subsubsection{Semantic-Aware Shape Conditioning \label{sec:condition}}
In contrast to image-conditioned 3D generation, we take the point cloud as input. To incorporate holistic shape information, we directly employ the VAE encoder to encode the input point cloud $\bm{X}$, which serves as the object-level condition $\bm{f}_o$. 

For the controllability over the decomposition process, we design the bounding box-based part-level cues extraction module, shown in \Cref{fig:pipeline}, to extract the part-level condition $\bm{f}_p$. First, we sample points  $\bm{X}_{\text{inbox}}$ within the given bounding box from the object point cloud. 
$\bm{X}_{\text{inbox}}$ is then encoded by a learnable encoder to form the part-level condition $\bm{f}_p$. In addition, to further improve the robustness to bounding box perturbations during inference, we apply augmentations involving random translations and moderate scaling to the bounding boxes during training. 

Furthermore, to facilitate more accurate decomposition of the part geometry, we enhance input conditions by concatenating point-wise semantic features encoded by $\text{P}^3\text{-SAM}$~\cite{P3_SAM} with the shape tokens encoded by the VAE encoder. The enhanced object-level and part-level conditional features, $\bm{f^{'}}_o$ and $\bm{f^{'}}_p$, are defined as:
\begin{equation}
\begin{split}
    \bm{f^{'}}_o & = \text{Concat}(\bm{f}_o, \mathcal{E}_{sem}(\bm{X})), \bm{f}_o = \mathcal{E}_o(\bm{X}) \\
    \bm{f^{'}}_p & = \text{Concat}(\bm{f}_p, \mathcal{E}_{sem}(\bm{X}_{\text{inbox}})), \bm{f}_p = \mathcal{E}_p(\bm{X}_{\text{inbox}})
\end{split}
\end{equation}
where $\mathcal{E}_o$ denotes the object-level VAE encoder, and $\mathcal{E}_p$ represents the learnable encoder in part-level cues extraction module. $\mathcal{E}_{sem}$ represents for the semantic encoder proposed in $\text{P}^3\text{-SAM}$. 
Note that to align with the shape tokens, the semantic feature is obtained by interpolated using the down-sampled XYZ positions from the shape encoder output, c.f. Figure~\ref{fig:pipeline}.
To enhance the robustness to the high-dimensional semantic feature, we randomly mask the semantic feature for partial points. 

Note that, when extracting the part-level condition, points belonging to other adjacent parts may be involved in the box of a certain part. However, thanks to the point-wise semantic feature and inter-part attention (discussed later in \Cref{sec:sync_part_gen}), different parts could help each other exclude points that do not belong to themselves. 

\subsubsection{Synchronized Part Generation}\label{sec:sync_part_gen}
Our proposed $\mathcal{X}$-Part simultaneously generates latent tokens for all parts of the whole object $\bm{O} =\{\bm{z_i}\}_1^K \in \mathbb{R}^{nK\times C}$ , where the object consists of $K$ parts and each part represented by $n$ latent tokens denoted as $\bm{z_i}$. Specifically, $\mathcal{X}$-part contains several DiT blocks and each block consists of one self-attention layer followed by two cross-attention layers (see \Cref{fig:pipeline}). 

Initially, self-attention is conducted within each part, providing intra-part awareness. However, this leads to the performance degradation at the boundaries between parts. Therefore, to enhance inter-part awareness, we extend the receptive field in half of the self-attention layers to encompass all part tokens, a design choice aligned with~\cite{lin2025partcrafter}.
\begin{equation}
\begin{split}
    &  \textbf{Attn}_{intra} = \textbf{softmax}(\frac{\sigma_q(\bm{z_i})\sigma_k(\bm{z_i})^T}{\sqrt{d}})\sigma_v(\bm{z_i})
    \\
    & \textbf{Attn}_{inter} = \textbf{softmax}(\frac{\sigma_q(\bm{z_i})\sigma_k(\bm{O})^T}{\sqrt{d}})\sigma_v(\bm{O})
\end{split}
\end{equation}
where $\sigma_{\{q|k|v\}}$ represent projection layers for the query, key and value. $d$ denotes the hidden dim for attention tokens.

Then, as shown in \Cref{fig:pipeline}, we inject the geometric condition, $\bm{f^{'}}_o$ and $\bm{f^{'}}_p$, by two layers of cross-attention module to improve the structural consistency of decomposition and preserve geometric details of the input object.

Furthermore, to enhance the distinctiveness between latent tokens of different parts, we incorporate a learnable part embedding for each individual part. Specifically, we initialize a codebook $\bm{E}\in \mathbb{R}^{l\times C}$ and assign a unique embedding to each part during training. Note that, to enable the decomposition of objects that contains more parts than the maximum limit for a single object in the dataset, we set $l$ a much larger number, and randomly select the unique embedding to each part.

\noindent
\textbf{Training.}
Building upon the conditioning framework established in \Cref{sec:condition}, we train the model using a flow matching objective~\cite{lipman2022flow} to transport noisy part tokens toward the target data distribution \( \bm{z}_0 \). Specifically, during the forward process, Gaussian noise $\bm{\varepsilon} \sim \mathcal{N}(0, \mathbf{I})$ is added to the data \( \bm{z}_0 \) according to a noise level \( t \), resulting in \( \bm{z}_t = t \bm{z}_0 + (1 - t) \bm{\varepsilon} \). The model is trained to predict the velocity field $\bm{v}=\bm{\varepsilon}-\bm{z}_0$that moves \( \bm{z}_t \) back toward \( \bm{z}_0 \), conditioned on both the object-level condition \( \bm{f^{'}}_o \) and the part-level condition \( \bm{f^{'}}_p \). The training objective is formulated as follows:
\begin{equation}
\mathcal{L}=\mathbb{E}_{\bm{z},t,\bm{\varepsilon}}\left[\left\Vert
(\bm{\varepsilon}-\bm{z}_0) - \bm{v}_{\theta}(\bm{z}_t,t,\bm{f^{'}}_o ,\bm{f^{'}}_p)
\right\Vert^2\right]
\end{equation}
where \( \bm{v}_{\theta} \) denotes the DiT-based neural network.
Given that the geometric complexity of an individual part is substantially lower than that of a complete object, we assign a reduced number of tokens to each part during both the VAE fine-tuning process and $\mathcal{X}$-Part training process. It significantly accelerates both training and inference while remaining the performance.

\subsection{Part Editing\label{sec:edit_pipeline}}
Leveraging the controllability and ease of manipulation provided by the bounding box, we further design a part-level editing pipeline for interactive part generation. Following Repaint~\cite{lugmayr2022repaint}, we adopt a training-free method to achieve two kinds of editing types, that is, split and adjust. The split operation refers to splitting the bounding box and generating several parts accordingly. The adjust operation means adjusting a certain bounding box so that the part and adjacent parts would be re-generated accordingly. Specifically, for parts indicated by the bounding box, their latent tokens are resampled and denoised while keeping tokens of other parts unchanged. 

\section{Experiments}
We begin by detailing the implementation of our model, including the network architecture, hyperparameters, and training procedure. We then conduct comparative evaluations against existing part generation methods. Furthermore, we perform ablation studies to validate the design choices of our framework. Finally, we demonstrate various downstream applications enabled by our method.
\subsection{Implementation Details}

\textbf{Network Architecture}
The DiT module consists of 21 DiT blocks, where skip connections are implemented by concatenating latent features along the channel dimension. During training, the number of tokens per part is set to 512, consistent with the VAE fine-tuning configuration. The self-attention layers at odd indices are configured to perform inter-part attention, thereby enhancing awareness of other parts. For the cross-attention modules, both the object condition and the part condition are represented with 2,048 tokens, providing detailed guidance for the generation process. The part embedding codebook contains 50 entries, and a unique embedding is randomly assigned to each part latent during both training and inference. In addition, we employ a Mixture-of-Experts (MoE) model for the linear output layers of the first six network blocks to efficiently enhance the learning capacity in the latent space.

\textbf{Training}
Our model is initialized from a pre-trained object generator, with its self-attention parameters loaded as the starting point. We use the Adam optimizer with a learning rate of $1e-4$ and apply gradient clipping with a maximum norm of $1.0$ to enhance training stability. The model was trained for approximately four days on 128 H20 GPUs. To further improve robustness, we randomly drop semantic features with a probability of $0.3$, and independently apply a $0.1$ dropout probability to the object condition, the part condition, or both during training. Additionally, we apply data augmentation to the bounding boxes by introducing random translations sampled from a uniform distribution $\mathcal{U}(-0.05, 0.05)$ and scaling factors sampled from the interval $[0.9, 1.1]$.

\textbf{Dataset Curation}
We use the part dataset introduced in P$^3$-SAM~\cite{P3_SAM}, which contains nearly 2.3 million objects with ground truth part segmentation. To create training pairs, each part of an object, as well as the object itself, is remeshed into a watertight mesh. A dataset of this scale significantly enhanced the generalizability of our diffusion-based shape decomposition method.

\subsection{Comparison}
Existing methods can be broadly categorized into two groups: 3D Shape Decomposition and Image-to-3D Part Generation. We compare our approach against representative methods from both categories at two distinct levels: the overall object level and the decomposed part level. This two-tier evaluation comprehensively validates our model's capacity for part decomposition and high-fidelity geometric generation.

\textbf{Evaluation Protocol}
We evaluate our method on 200 samples from the ObjaversePart-Tiny dataset, each comprising rendered images and corresponding ground-truth part geometries. To assess geometric quality, we employ Chamfer Distance (CD) and F-Score. The F-Score is computed at two different thresholds $[0.1, 0.5]$ to capture both coarse-level and fine-level geometric alignment. Prior to metric computation, each object is normalized to the range $[-1, 1]$. To ensure pose-agnostic evaluation, we rotate each object by $[0, 90, 180, 270]$ degrees and report the best score among these orientations as the final metric.

\begin{table}[]
\centering
 \scalebox{0.8}{
\begin{tabular}{lccccc}
\toprule
\textbf{Method} & CD↓ & Fscore-0.1↑ & Fscore-0.05↑ \\
\midrule
SAMPart3D  & 0.15 & 0.73 & 0.63 \\
PartField  & 0.17 & 0.68 & 0.57 \\
HoloPart  & 0.26 & 0.59 & 0.43  \\
OmniPart  & 0.23 & 0.63 & 0.46 \\
Ours      & \textbf{0.11} & \textbf{0.80} & \textbf{0.71}  \\
\bottomrule
\end{tabular}
}
\caption{Quantitative part decomposition results.}
\label{tab:part_level_metric}
\end{table}

\begin{figure*}[!ht]
  \centering
  \includegraphics[width=1\linewidth]{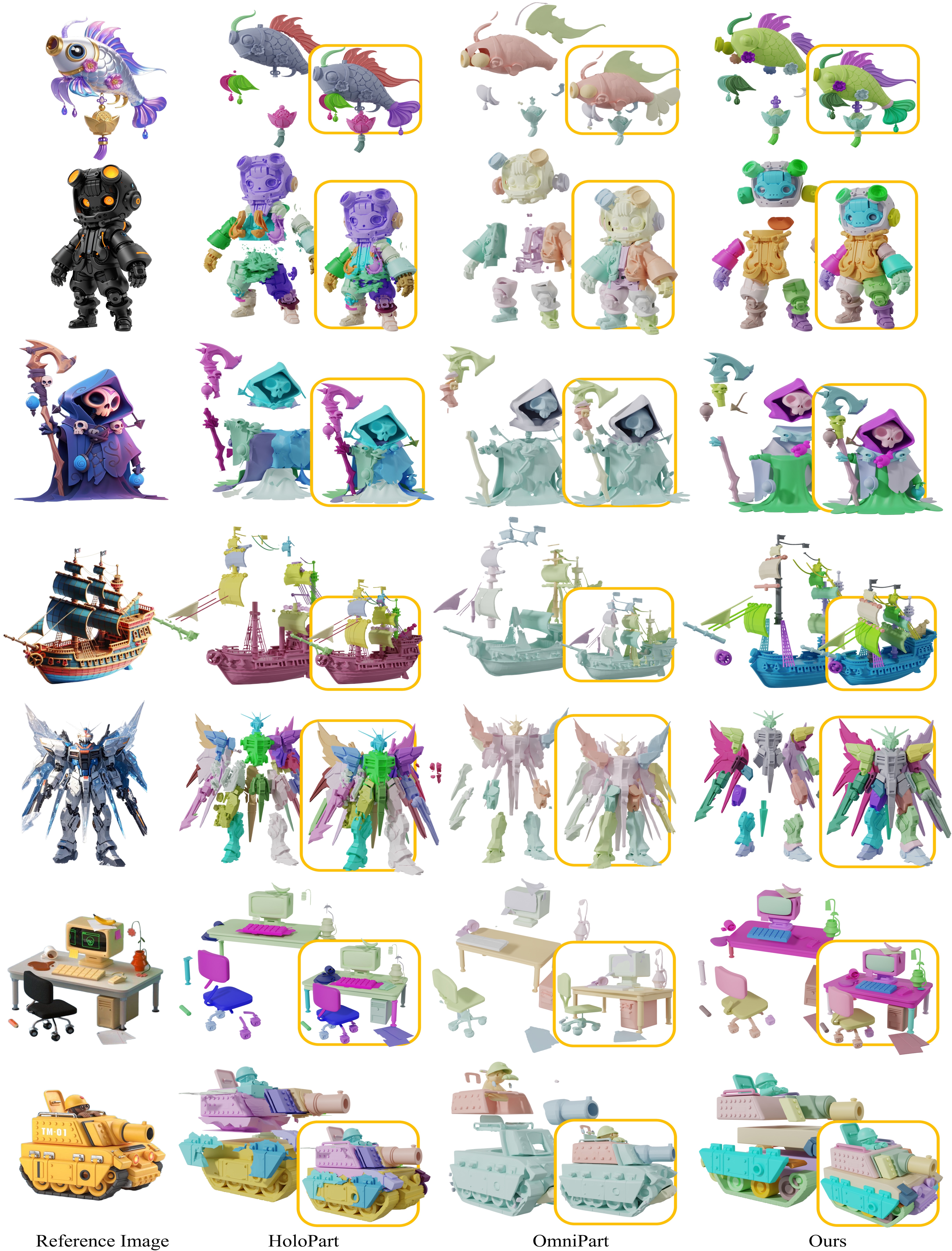}
  \caption{\textbf{Qualitative shape decomposition results.} Note the input shapes for HoloPart and Ours are obtained from Hunyuan3D-2.5~\cite{lai2025hunyuan3d}, while OmniPart leverage the shape from trellis~\cite{xiang2025structured}.}
  \label{fig:parts}
\end{figure*}

\textbf{3D Shape Decomposition.} This experiment aims to evaluate and compare the geometric decomposition capabilities of different methods, validating that our approach achieves a deeper structural understanding and decomposition of objects while generating higher-quality part geometries. Our method takes a ground-truth watertight surface as input and automatically generates decomposed parts; We compute metrics between the generated parts and the ground-truth parts.
We first compare against segmentation-based methods such as Sampart3D \cite{yang2024sampart3d} and PartField \cite{liu2025partfield}, which also take the same watertight mesh as input. The segmented results are directly compared to the ground truth parts. In addition, we include generative methods such as HoloPart \cite{yang2025holopart} and OmniPart \cite{yang2025omnipart}. HoloPart also uses the ground-truth watertight point cloud as input. Although OmniPart does not directly take a 3D shape as input, it first generates a coarse geometry and then performs part decomposition. To eliminate the influence of segmentation quality, we replace the Sampart3D segmentation used in HoloPart with our own trained segmentation model, and provide OmniPart with 2D part masks rendered from the ground-truth parts.
As shown in \Cref{tab:part_level_metric}, segmentation-based methods can decompose part points on the input watertight surface but fail to produce complete part geometries. Our method outperforms all baselines in decomposition quality, even when OmniPart is supplied with ground-truth 2D masks. Furthermore, as illustrated in \Cref{fig:parts}, our approach significantly surpasses other methods in the geometric quality of the generated parts.

\begin{table}[]
\centering
 \scalebox{0.8}{
\begin{tabular}{lccccc}
\toprule
\textbf{Method} & CD↓ & Fscore-0.1↑ & Fscore-0.05↑ \\
\midrule
Part123  & 0.42 & 0.36 & 0.20\\
HoloPart & 0.09 & 0.88 & 0.73 \\
PartCrafter  & 0.20 & 0.66 & 0.45 \\
PartPacker  & 0.11 & 0.85 & 0.65 \\
OmniPart  & \textbf{0.08} & 0.91 & 0.77 \\
Ours      & \textbf{0.08} & \textbf{0.92} & \textbf{0.78} \\
\bottomrule
\end{tabular}
}
\caption{Quantitative holistic shape generation results.}
\label{tab:parts}
\end{table}
\begin{figure*}[!ht]
  \centering
  \includegraphics[width=1\linewidth]{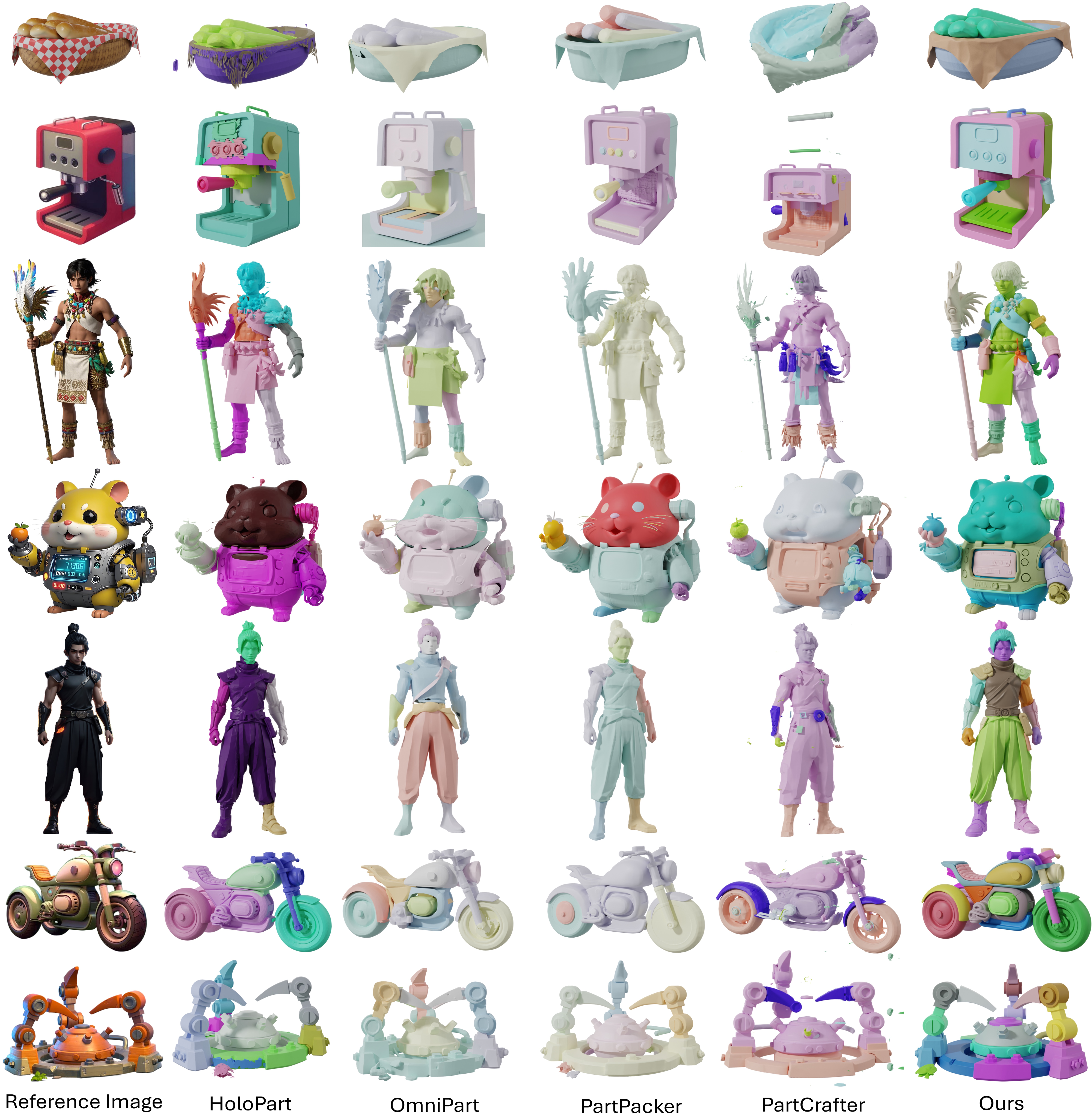}
  \caption{\textbf{Qualitative shape decomposition results.} Note the input shapes for HoloPart and Ours are obtained from Hunyuan3D-2.5, OmniPart leverage the shape from Trellis, PartCrafter and PartPacker are image to 3D methods, they do not rely on shapes.}
  \label{fig:overall}
\end{figure*}

\textbf{Image-to-3D Part Generation.} Leveraging existing image-to-3D generative models, we extend our method to the task of image-to-3D part generation. Specifically, given a reference image, we first generate a watertight mesh using an off-the-shelf image-to-3D model \cite{zhang2024clay,lai2025hunyuan3d,li2025triposg}, which is then fed into our pipeline for decomposition into parts. Similar to the previous experiment, we compare our approach not only against HoloPart and OmniPart, but also against methods that directly generate parts from images, such as PartPacker \cite{tang2024partpacker}, PartCrafter\cite{lin2025partcrafter}, and Part123\cite{liu2024part123}. The input to OmniPart remains consistent with the setup above, while both HoloPart and our method use the same generated mesh as input.
Since different methods may produce divergent part structures, making it difficult to establish accurate correspondences with ground-truth parts. We compare only the overall object geometry composed of all generated parts. As shown in \Cref{tab:parts}, our method produces final objects with higher geometric quality and better alignment to the ground truth. \Cref{fig:parts} visually demonstrates the structural plausibility and high quality of our results. Moreover, our decomposition is more refined, often generating a larger number of semantically reasonable parts.

\subsection{Applications}

\begin{figure*}[!ht]
  \centering
  \includegraphics[width=1\linewidth]{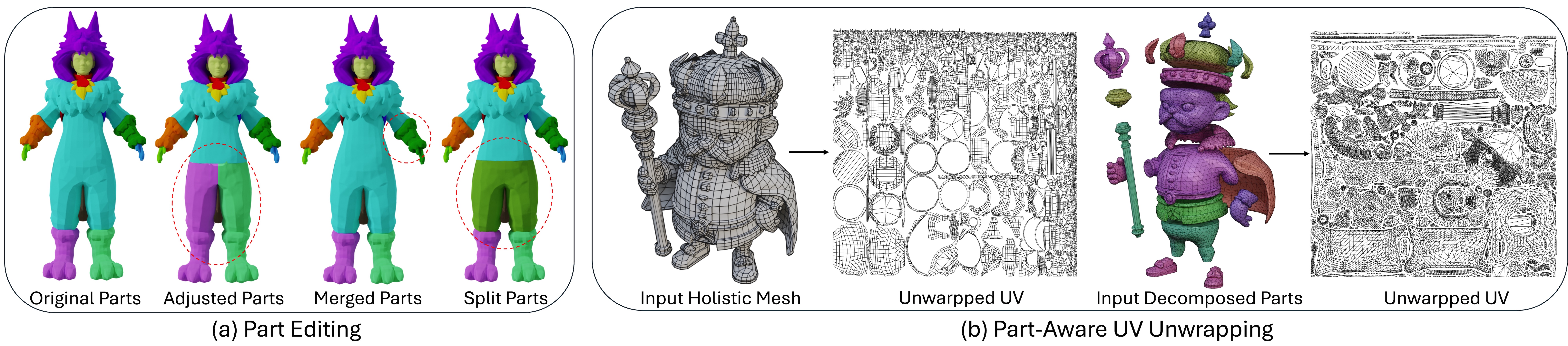}
  \caption{Demonstration of two representative applications of our method. Subfigure (a) shows the results of bounding box-controlled part generation, while subfigure (b) illustrates improved UV unwrapping performance achieved through part-based decomposition.}
  \label{fig:application}
\end{figure*}

\textbf{Part Editing}
Since our method directly takes point clouds and bounding boxes as input, it enables intuitive control over part decomposition and facilitates various part-level editing operations. As demonstrated in \Cref{fig:application}(a), users can easily adjust both the position and scale of bounding boxes to influence the geometric characteristics of the generated parts. Specifically, as described in \Cref{sec:edit_pipeline}, part generation can be controlled in multiple ways: modifying the location and size of a bounding box alters the shape and coverage of the corresponding part; merging adjacent bounding boxes results in the fusion of multiple parts into a single component; and splitting a bounding box leads to the decomposition of a part into finer structures.

\textbf{Part-Aware Un-wrapping.}
UV unwrapping is an essential step in 3D content creation pipelines. 
Fig.~\ref{fig:application} compares the UV maps generated by unwrapping a holistic mesh and part-decomposed meshes respectly.
Part-decomposed mesh are processed by unwrapping each of the part separatedly.
Decomposing shapes into part greatly simplify Un-wrapping process and makeing UV maps more compact and semantically meanningful.  

\subsection{Ablation Study}
\begin{table}[]
\centering
\begin{tabular}{l|ccc|ccc}
\toprule
\multirow{2}{*}{Method} & \multicolumn{3}{c|}{Part-level} & \multicolumn{3}{c}{Overall-level} \\
                        & CD $\downarrow$ & F1-0.1 $\uparrow$ & F1-0.05 $\uparrow$ & CD $\downarrow$ & F1-0.1 $\uparrow$ & F1-0.05 $\uparrow$ \\
\midrule
W/o part embedding  & 0.13 & 0.78  & 0.68 & 0.04 & 0.97 & 0.92 \\
W/o object-cond  & 0.12 & 0.79 & 0.70 & 0.03 & 0.97 & 0.93 \\
W/o part-cond  & 0.27 & 0.57 & 0.47 & 0.03 & \textbf{0.98} & 0.95 \\
W/o semantic-feat  & 0.12 & 0.78 & 0.69 & 0.04 & 0.97 & 0.92 \\
W/o inter-part self-attn  & 0.12 & 0.79  & 0.70 & 0.03 & 0.97 & 0.94  \\
Ours      & \textbf{0.11} & \textbf{0.80} & \textbf{0.71} & \textbf{0.02} & \textbf{0.98} & \textbf{0.96}  \\
\bottomrule
\end{tabular}
\caption{Based on the ground-truth bounding boxes, we compute part-level and object-level metrics for different modules on the ObjaversePart-Tiny dataset.}

\label{tab:ablation}
\end{table}

\begin{figure*}[!ht]
  \centering
  \includegraphics[width=1\linewidth]{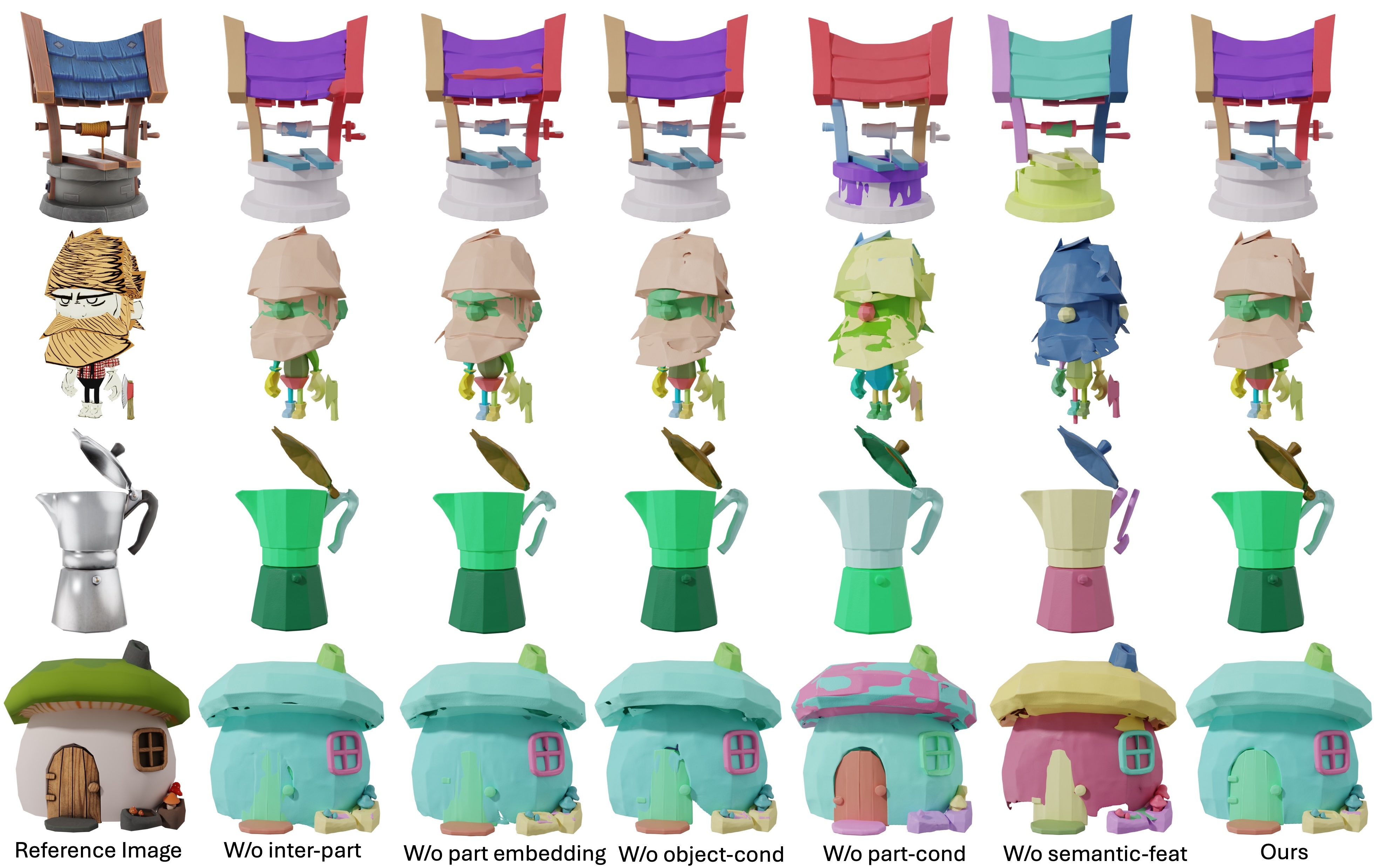}
  \caption{Part generation results under different module ablation settings.}
  \label{fig:ablation}
\end{figure*}

As shown in \Cref{tab:ablation}, we conduct a series of ablation studies to validate the effectiveness of each component in our proposed framework, all of which contribute to improved model performance. We analyze the roles of individual components in detail. The intra-part and inter-part attention mechanism enhances the representation of part-level latents while maintaining a global contextual view across all parts. The part embedding module introduces distinctiveness among the latent representations of different parts. The object-level condition provides priors about the overall geometry of the shape. Meanwhile, the part-level condition offers detailed information indicating coarse part location and scale. Additionally, the semantic point feature supplies semantic cues that facilitate structurally coherent shape decomposition. We further provide visualizations of representative results in \Cref{fig:ablation} to illustrate the impact of each component.

\section{Conclusion and Limitation}

\textbf{Conclusion}
We introduce $\mathcal{X}$-Part, a purely geometry-based part generation framework that takes bounding boxes as input to decompose complete 3D objects into structured parts. Compared to existing approaches, our method better preserves geometric quality and fidelity in the generated parts, while also offering easier integration into 3D content creation pipelines, thereby significantly reducing the complexity of downstream tasks. Additionally, our method allows users to alter part decomposition strategies by adjusting bounding boxes, thereby enabling more intuitive control and flexible editing. To enhance the model’s structural understanding, we incorporate semantic point features that provide high-level shape semantics. Our approach supports the generation of up to 50 distinct parts, which sufficiently covers most practical application scenarios.

\textbf{Limitation}
Our method currently relies on geometric cues for decomposition and lacks guidance from physical principles, which may limit its ability to meet certain application-specific decomposition requirements. Additionally, since the latent codes of all parts are processed simultaneously through the diffusion model, inference time increases with the number of parts, posing a challenge for real-time usage when handling high-part-count objects.

\bibliography{iclr2026_conference}
\bibliographystyle{iclr2026_conference}


\end{document}